# POVERTY, INCOME INEQUALITY AND GROWTH IN BANGLADESH: REVISITED KARL-MARX

**MD NIAZ MURSHED CHOWDHURY**
Department of Economics
University of Nevada Reno
mdniazc@nevada.unr.edu

**MD. MOBARAK HOSSAIN**
Department of Economics
University of Nevada Reno

## ABSTRACT

This study tries to find the relationship among poverty inequality and growth. It also tries to connect the Karl Marx's thoughts on functional income distribution and inequality in capitalism. Using the Household Income and Expenditure Survey of 2010 and 2016 this study attempt to figure out the relationship among them. In Bangladesh about 24.3% of the population is living under poverty lines and 12.3% of its population is living under the extreme poverty line. The major finding of this study is poverty has reduced significantly from 2000 to 2016, which is more than 100% but in recent time poverty reduction has slowed down. Despite the accelerating economic growth, the income inequality also increasing where the rate of urban inequality exceed the rural income inequality. Slower and unequal household consumption growth makes sloth the rate of poverty reduction. Average annual consumption fell from 1.8% to 1.4% from 2010 to 2016 and poorer households experienced slower consumption growth compared to richer households.

# I. INTRODUCTION

Global inequality is the major concern in the recent era, which is rising at an alarming rate. According to recent Oxfam report, 82% of the wealth generated last year went to the richest 1% of the global population, which the 3.7% billion people who make up the poorest half of the world population had no increase in their wealth in another word poorest half of the world got nothing. Which is the clear indication of exploitation of labor and these economics rewards are increasingly concentrated at the top from bottom. Top 10% of South African Society receives 50% of all wage income while the bottom 50% of the workforce receives only 12% of all wages. A CEO in the US earns the same as the ordinary worker makes during the whole year and on average, it takes just over four days for a CEO from the top five companies in the garment sector to earn what an ordinary Bangladeshi woman worker earns in her whole lifetime.

Bangladesh has achieved a high growth rate in recent decades, during which its annual gross domestic product (GDP) growth rate is 7.0 percent and expected to be 7.2 percent by the end of the quarter. It is projected that GDP annual Growth rate is around 7.00 percent in 2020, according to Trading Economics global macro models and analysts expectations. The major challenge throughout the globe is to reduce or control economic inequality and this economic inequality arises through the distribution of income, consumption, wealth or assets. Household level information of Bangladesh suggests that the distribution of income is much more unequal than the distribution of consumption. In Bangladesh top 5% income people has taken over of 95% of total income, which indicates an uneven distribution of wealth. Income inequity is measured by the GINI coefficient and it has been substantially used as a measure of income inequality in the last few decades. Bangladesh is one of the most promising economies in the 21$^{st}$ century but the incidence of poverty and income inequality is very high among Asian countries. This high economic growth is not adequately benefiting a large part of the people because of the rising inequality where the GINI index increased by 0.93% from 2010 to 2016, which clearly indicates inequality rises in Bangladesh.

**Figure-1: TOP 10 FASTEST-GROWING UHNW COUNTRIES 1**

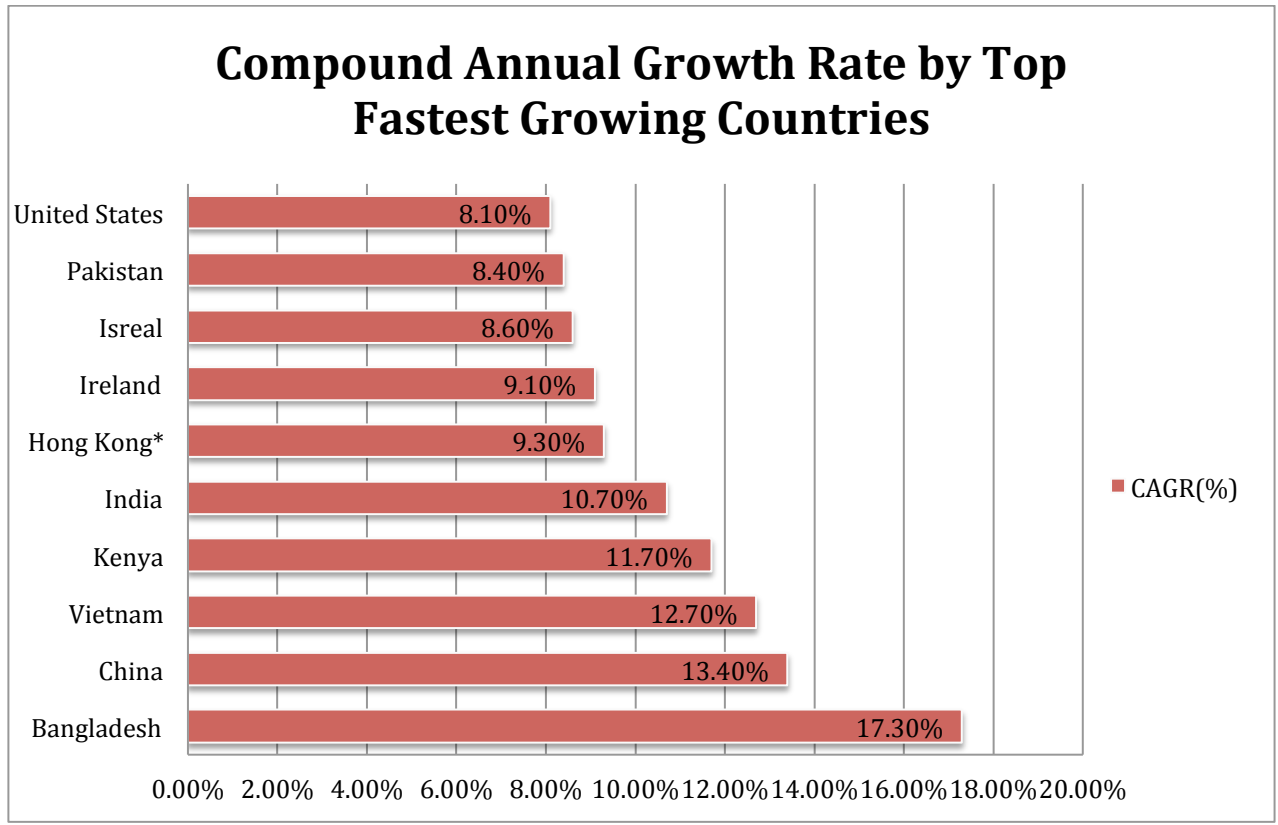

Hong Kong is a semi-autonomous, special character region in China
CAGR stands for compound annual growth rate
Source: Wealth-X

This report shows that rich in Bangladesh rising faster than anywhere in the globe. Bangladesh has been ranked the fastest growing country with an increasing number of rich population in the world, according to World Ultra Wealth-X. The number of ultra high net worth (UHNW)[1] individuals in Bangladesh rose by 17.3 percent during the period where the United States rose by 8.1%.

---

[1] UHNW individuals are defined as people with invested assets of at least $30 million, usually excluding personal assets and property such as a primary residence,bles and consumer durables.

Inequality in capitalism is the main concern in the recent era. Karl Marx wrote extensively about capitalism and his strong arguments against the traditional economic system is about the inevitability of inequality. Capitalism is a broken system in this world, it concentrates wealth towards up from the bottom and creates situations that stimulate inequality between the rich and the poor or more appropriately, between the capitalists and the working class. Capitalist exploits their laborer not giving the proper share of profit using the concept of property rights and exploitation become legal and morally acceptable. Capitalism is an economic system based on the private, which allows individuals to own and operate their own business while also promoting open competition and the free market. The objective of this study is to find the relationship among poverty, inequality, and growth. It also tries to connect the Karl Marx's thoughts on functional income distribution and inequality in capitalism.

## II. METHODOLOGY

This study has used the data of Household Income and Expenditure Survey (HIES) of 2000, 2005, 2010 and 2015. These are the secondary source of data conducted by the Bangladesh Bureau of Statistics (BBS). I also used different data sources like World Bank Data Bank, IMF, OECD, Bangladesh Bank, Ministry of Finance and so on. In addition, various statistical report, for example, OXFAM, book and different published articles were reviewed to write the paper. Different statistical techniques, graph, chat, line chart, pie chart, bar diagram used to analyze this data.

## III. POVERTY, INEQUALITY, AND GROWTH

Bangladesh has been suffering in extreme poverty last few decades. Bangladesh has achieved impressive progress to alleviate poverty in the last decade and the government of Bangladesh has taken numerous steps to fight against poverty. The poverty rate for Bangladesh in 2016 was 24.3% that is about 101% less than the year of 2000 and the extreme poverty rate for 2016 was 12.9% that's is 165% lower than 2000. The international extreme poverty rate was 13.8 percent (18.5% in 2010) in the same year, which is higher than Bangladesh extreme poverty rate. Between 2010 and 2016 poverty fell significantly but this rate has slowed down in recent years. Poverty fell faster rate in rural areas than in urban areas and rural poverty rates declined from 35.2 to 26.4 percent, while urban poverty rates decreased from 21.3 to 18.9 percent. The pace of economic growth is outstanding since 2010 with a GDP growth rate on average 6.8% per year between the periods of 2010 to 2018. Despite the high, the rate of poverty reduction is comparatively slow, which in the indication of the unequal distribution of income. Unequal household consumption growth of Bangladesh is the main cause of slower rate of poverty reduction. Accelerating economic growth could not able to reduce poverty at time same pace.

**Figure-2: Poverty versus extreme poverty in Bangladesh.**

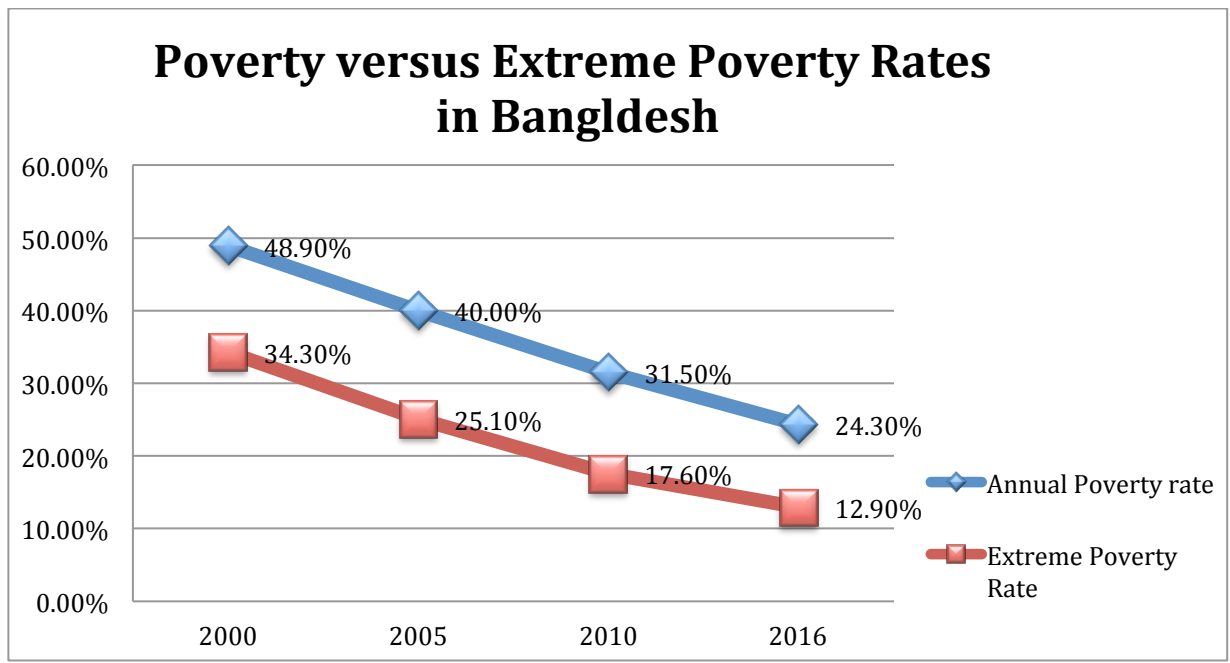

## IV. TREND IN POVERTY- NATIONAL, URBAN AND RURAL

National poverty (upper poverty line)) headcount rate has declined by 101 percent (or 24.6 percentage point) between 2000 and 2016. In 2016, almost 1 in 4 (24.3%) Bangladeshi live in under poverty line, while more rural population than urban live under poverty thought rural poverty rates (33.3%) decline faster rate than urban poverty rates (29.6%).

**Table 1: Poverty Headcount Rates (%)**

|  | Upper Poverty Line | | | | Lower Poverty Line | | | |
| --- | --- | --- | --- | --- | --- | --- | --- | --- |
|  | 2000 | 2005 | 2010 | 2016 | 2000 | 2005 | 2010 | 2016 |
| **National** | 48.9 | 40 | 35 | 24.3 | 34.3 | 25.1 | 17.6 | 12.9 |
| **Urban** | 35.2 | 28.4 | 24.5 | 18.9 | 19.9 | 14.9 | 12.4 | 10.8 |
| **Rural** | 52.3 | 43.8 | 35.2 | 26.4 | 37.9 | 28.6 | 23.4 | 19.8 |

Source: HIES 2000, 2005 and 2016; using poverty line estimated with HIES (2010) and deflated to adjust for inflation during 2000-16

A notable feature of poverty reduction between 2000 and 2016 is that there is a downward trend in poverty reduction in all the three levels like national, rural and urban. The below line chart and diagram shows that the rural poverty line always higher than national and urban poverty rates through rural poverty rates decrease faster pace. If we consider the lower or extreme poverty rates in the same period of time, rural poverty was always higher than national and urban poverty. Bangladesh has gained tremendous success to fight against extreme poverty, which went down to 12.9% in 2016 from 34.3% (2000).

**Figure-3: Comparison among National, Urban and Rural Poverty Line**

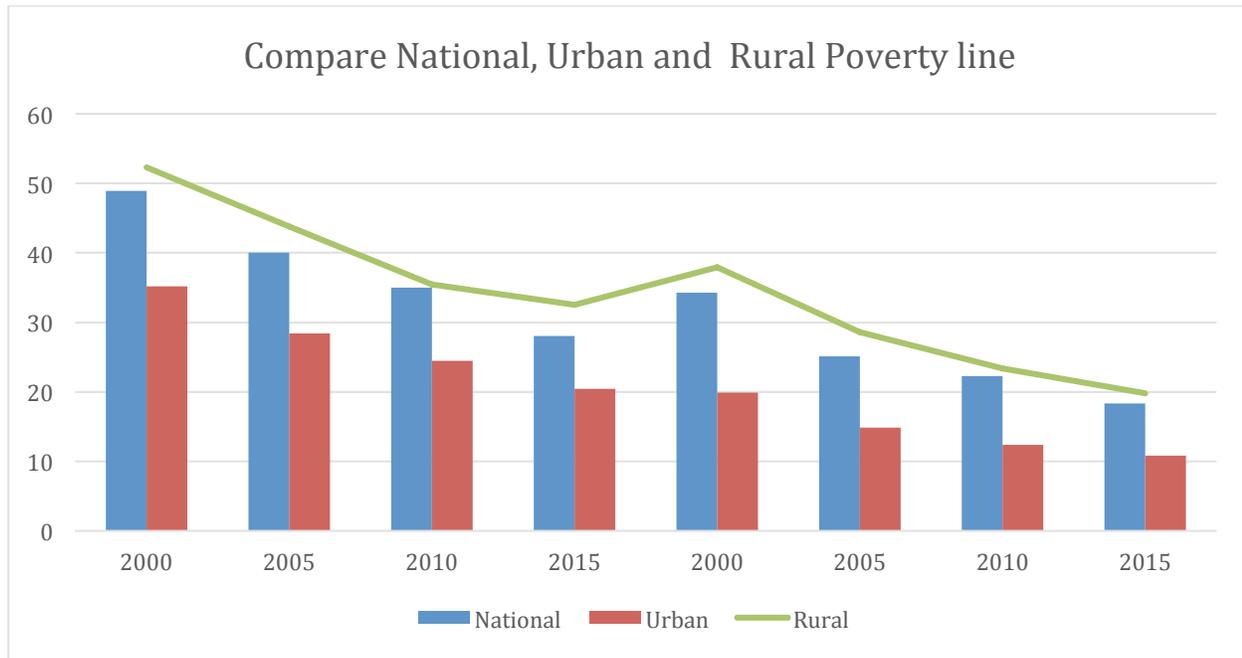

## V. INCOME SHARE OF HOUSEHOLDS IN QUINTILES

Table 2 and Graph 2 shows the income share of different households between the periods of 1973 to 2016. I observed an overall downward trend in $1^{st}$ to $4^{th}$ quintile but income share has increased in $5^{th}$ quintile or top quintile. $1^{st}$ quintile decreased to 5.26 percent from 7.20 percent over the period of time (decreased by 36%). Similarly, $2^{nd}$ quintile decreased by an annual average rate of 14.9%, $3^{rd}$ quintile decreased by 8.7% and $4^{th}$ quintile declined by 7.3%. It indicates that poorer household suffers in terms of share in income, in other words, the poorer the household the more they suffer. In contrast, there was an upward in income share in $5^{th}$ quintile, $9^{th}$ decile, and $10^{th}$ decile. The gain in the income share of the households in the top quintile, the gain was accounted 6.03% in the top quintile. We also observed that there was almost no gain in $9^{th}$ quintile/decile but in $10^{th}$ quintile/decile the gain in income share was about 6.27%.

**Table 2: Income (Share percent) accruing to Household (Quintile/Decile) National 1973-2016**

| Year | 1st Quintile | 2nd Quintile | 3rd Quintile | 4th Quintile | 5th Quintile | 9th Decile | 10th Decile |
|---|---|---|---|---|---|---|---|
| **1973-74** | 7.00 | 11.30 | 15.10 | 22.80 | 44.40 | 16.00 | 28.40 |
| **1981-82** | 6.64 | 10.72 | 15.20 | 22.12 | 45.32 | 15.79 | 29.53 |
| **1983-84** | 7.2 | 11.75 | 15.94 | 21.73 | 43.38 | 15.08 | 28.30 |
| **1985-86** | 6.99 | 11.18 | 15.07 | 20.7 | 48.04 | 14.58 | 31.46 |
| **1988-89** | 6.64 | 10.89 | 15.05 | 21.23 | 46.20 | 15.2 | 31.00 |
| **1991-92** | 6.52 | 10.89 | 15.53 | 22.19 | 44.96 | 15.64 | 29.32 |
| **1995-96** | 5.71 | 9.83 | 13.88 | 20.5 | 50.08 | 15.40 | 34.68 |
| 2000 | 6.15 | 9.68 | 13.17 | 18.79 | 52.01 | 14.00 | 38.01 |
| 2005 | 5.26 | 9.10 | 13.13 | 19.79 | 52.71 | 15.07 | 37.64 |
| 2010 | 5.22 | 9.10 | 13.33 | 20.56 | 51.79 | 15.94 | 35.85 |
| 2016 | 5.12 | 8.90 | 13.27 | 20.34 | 50.43 | 16.07 | 34.67 |
| Change during 1973-2016 | -1.88 | -2.4 | -1.83 | -2.46 | 6.03 | 0.07 | 6.27 |
| Annual Rate Change | -0.72 | -0.56 | -0.34 | -0.29 | 0.49 | 0.01 | 0.74 |

**Figure-4: Income (Share percent) accruing to Household (Quintile/Decile) National 1973-2016**

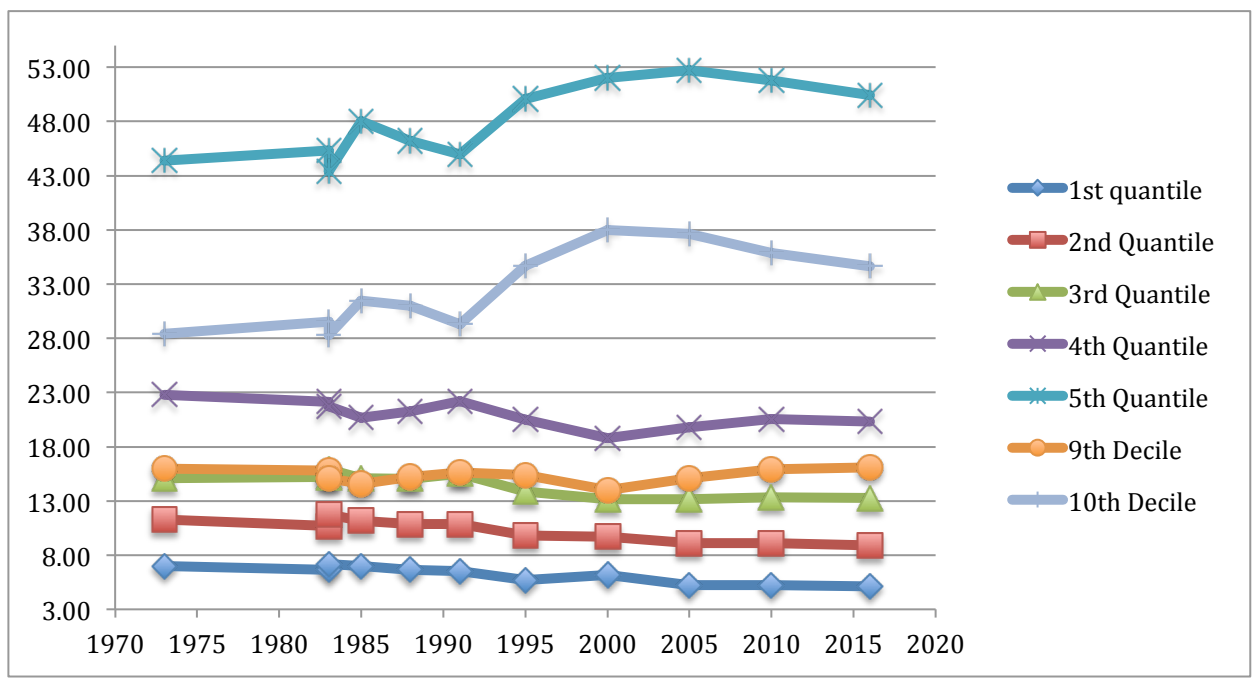

**VI. RATIO OF INCOME SHARE IN TOP 10% AND BOTTOM 10%**

This ratio (top 10% divided by bottom 10%) has been used to measure of saturation of income inequality. From the table and graph below, at the national level this ratio has increased by 9.1 percentage point from 10.14% in 1973 to 19.75% in 2016. This result shows that the bottom 10% of the country's population has been going through a very bad situation in terms of households income share. For the rural area increased by 6.56 percentage point, while in an urban area increased by 10.55 percentage point, which is significantly higher than in rural areas. This income inequality increased by 2.61% annual average rate that is substantially higher than the rural areas (increased by the annual average rate of 1.15%). Based on the discussion, we observed a very high level of income inequality in urban areas of Bangladesh for the years of 2000 and 2016.

**Table 3: Ratio of Income Share Top 10% and bottom 10%**

| Year | Ratio=Top10% /Bottom 10% | | |
|---|---|---|---|
| **Year** | National | Rural | Urban |
| **1973** | 10.14 | 11 | 9.12 |
| **1981** | 10.7 | 9.4 | 10.85 |
| **1983** | 9.79 | 9.45 | 9.87 |
| **1985** | 11.19 | 10.62 | 10 |
| **1988** | 11.74 | 10.98 | 11.3 |
| **1991** | 11.33 | 10.5 | 11.53 |
| **1995** | 15.48 | 11.81 | 18.78 |
| **2000** | 15.77 | 11.72 | 20.45 |
| **2005** | 18.82 | 15.07 | 22.82 |
| **2010** | 17.92 | 15.2 | 17.54 |
| **2016** | 19.75 | 17.56 | 19.67 |
| **Change During 1973-2016** | 9.61 | 6.56 | 10.55 |
| **Average Annual Rate of Change (%)** | 2.24 | 1.15 | 2.61 |

**Figure-5: Ratio of Income Share Top 10% and bottom 10%**

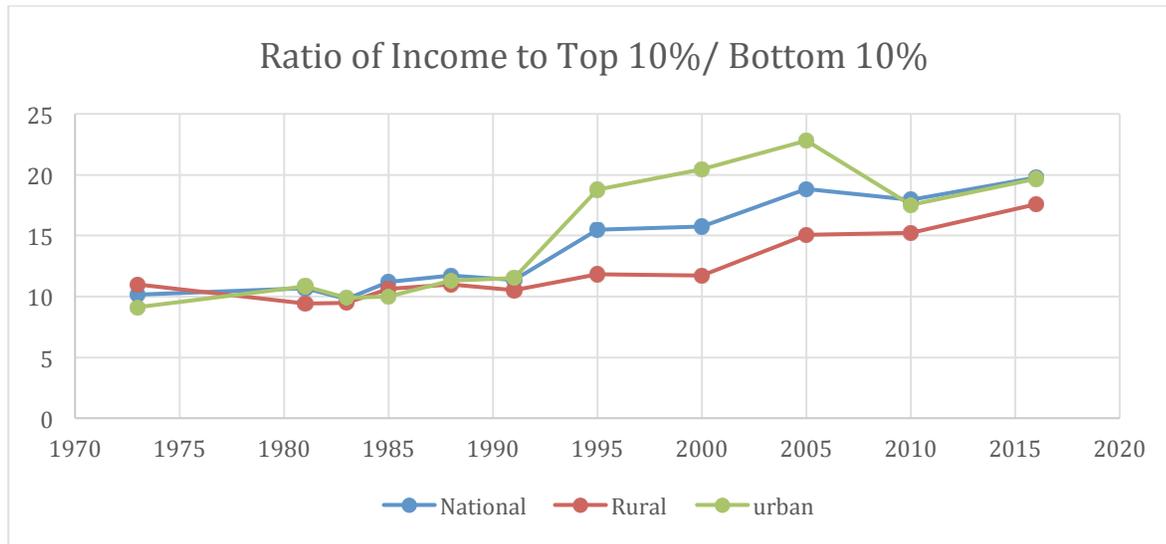

## VII. GINI INDEX FOR INCOME

GINI coefficient or GINI index has been used to measure the income inequality and it has risen significantly last few decades throughout the world. The value of GINI coefficient varies from 0 to 1, 0 meaning there is no inequality (perfect equality) and 1 indicates perfect inequality where income is highly skewed to an individual (posses all the income). The values of GINI coefficient was higher to urban areas from 2000 to 2016 compared to rural areas, which indicates that income inequality higher in an urban area than rural areas. The data shows that income equality has been rising in Bangladesh over the period.

**Table-4: GINI index for income**

| Year | National | Rural | Urban |
|---|---|---|---|
| **1973** | 0.36 | 0.35 | 0.38 |
| **1981** | 0.39 | 0.36 | 0.41 |
| **1983** | 0.36 | 0.35 | 0.37 |
| **1985** | 0.38 | 0.36 | 0.37 |
| **1988** | 0.38 | 0.37 | 0.38 |
| **1991** | 0.39 | 0.36 | 0.4 |
| **1995** | 0.43 | 0.38 | 0.44 |
| **2000** | 0.45 | 0.39 | 0.5 |
| **2005** | 0.47 | 0.43 | 0.5 |
| **2010** | 0.46 | 0.43 | 0.45 |
| **2016** | 0.48 | 0.44 | 0.43 |
| **Change During 1973 to 2016** | 0.12 | 0.09 | 0.05 |
| **Average Annual rate of change** | 0.76 | 0.62 | 0.5 |

**Figure-6: GINI index for income**

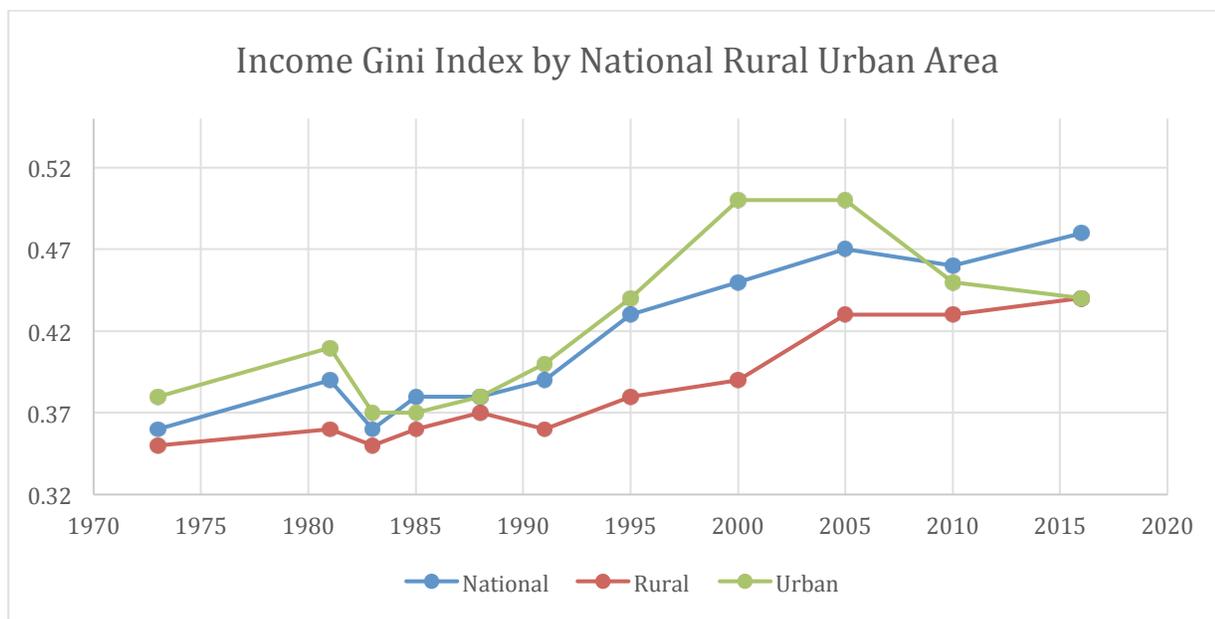

## VIII. GINI COEFFICIENT OF PER CAPITA INCOME

When we consider GINI Coefficient on per capita income for the periods of 2000 to 2016, we observed the similar results that we obtained for household income. We observed that from the below diagram that rural income inequality has increased over the periods but there was a fall in urban income inequality between the periods of 2005 to 2010 and then it increased.

**Table-4: GINI Coefficient of Per Capita Income**

| Year | National | Rural | Urban |
|---|---|---|---|
| 2000 | 0.451 | 0.393 | 0.497 |
| 2005 | 0.467 | 0.428 | 0.497 |
| 2010 | 0.458 | 0.431 | 0.452 |
| 2016 | 0.461 | 0.434 | 0.456 |
| Change During 2000 and 2016 | 0.01 | 0.031 | 0.041 |
| Average Annual Rate of change (percent) | 0.16 | 0.98 | -0.91 |

**Figure-7: GINI Coefficient of Per Capita Income**

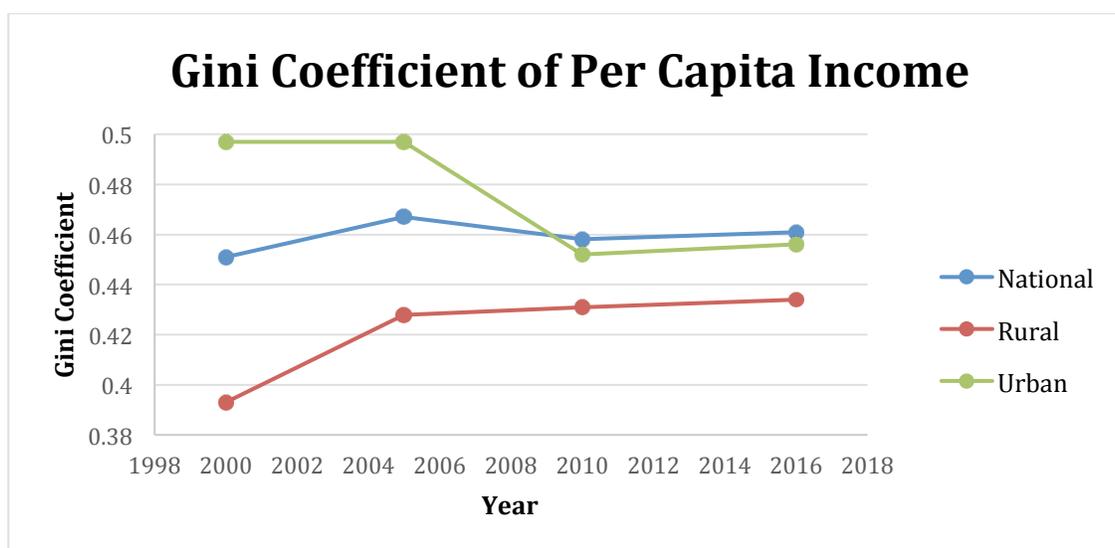

**IX. INCOME SHARE TO BOTTOM 40% OF HOUSEHOLDS**

Income share to bottom 40% of households decreased from 18.3% in 1973 to 14.28% in 2016, which is accounted an overall decrease of a 4.02-percentage point with the annual average rate of decrease has been 0.60 percent. We see from the below diagram that there is a downward trend in income share to the bottom 40% of households over the periods.

**Table-5: Income Share (Percent) Accruing to Bottom 40 Percent Households**

| Year | Income Share (Percent) Accruing to Bottom 40 Percent Households |
|---|---|
| **1973** | 18.3 |
| **1981** | 17.36 |
| **1983** | 18.95 |
| **1984** | 18.17 |
| **1988** | 17.53 |
| **1991** | 17.41 |
| **1995** | 15.54 |
| **2000** | 15.83 |
| **2005** | 14.36 |
| **2010** | 14.32 |
| **2016** | 14.28 |
| **Change From 1973 to 2016** | -4.02 |
| **Average Annual Rate of Change (Percent)** | -0.63 |

**Figure-8: Income Share (Percent) Accruing to Bottom 40 Percent Households**

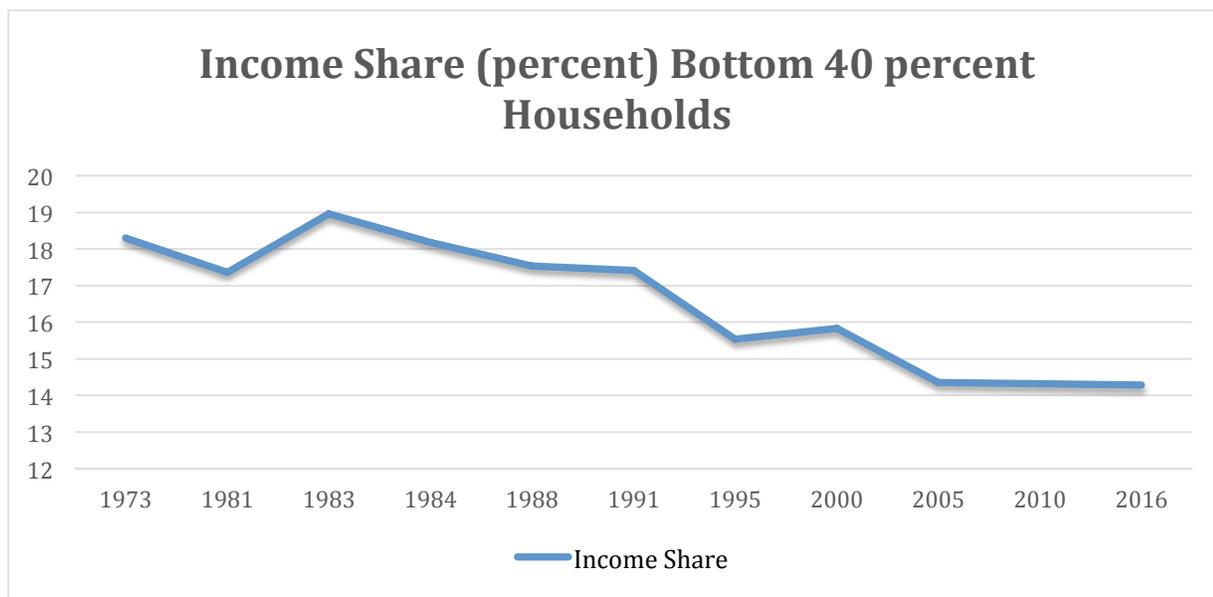

## X. RELATIONSHIP BETWEEN INEQUALITY AND PER CAPITA INCOME

Simon Kuznets first observed the pattern distribution of income, he said in the early stage of development both growth and inequality shall go together. In Bangladesh, both per capita income and GINI coefficient increases is the same pace, which means growth accelerate income inequality in Bangladesh.

**Table-6: Relationship between inequality and per capita**

| Year | Per Capita Income (Thousand TK) | GINI index |
|---|---|---|
| **1973** | 9.9 | 0.36 |
| **1981** | 10.3 | 0.39 |
| **1983** | 10.8 | 0.36 |
| **1985** | 11.1 | 0.38 |
| **1988** | 11.7 | 0.38 |
| **1991** | 12.4 | 0.39 |
| **1995** | 13.9 | 0.43 |
| **2000** | 16.6 | 0.45 |
| **2005** | 20.5 | 0.47 |
| **2010** | 27.1 | 0.46 |
| **2016** | 36.8 | 0.48 |

**Figure-9: Relationship between inequality and per capita**

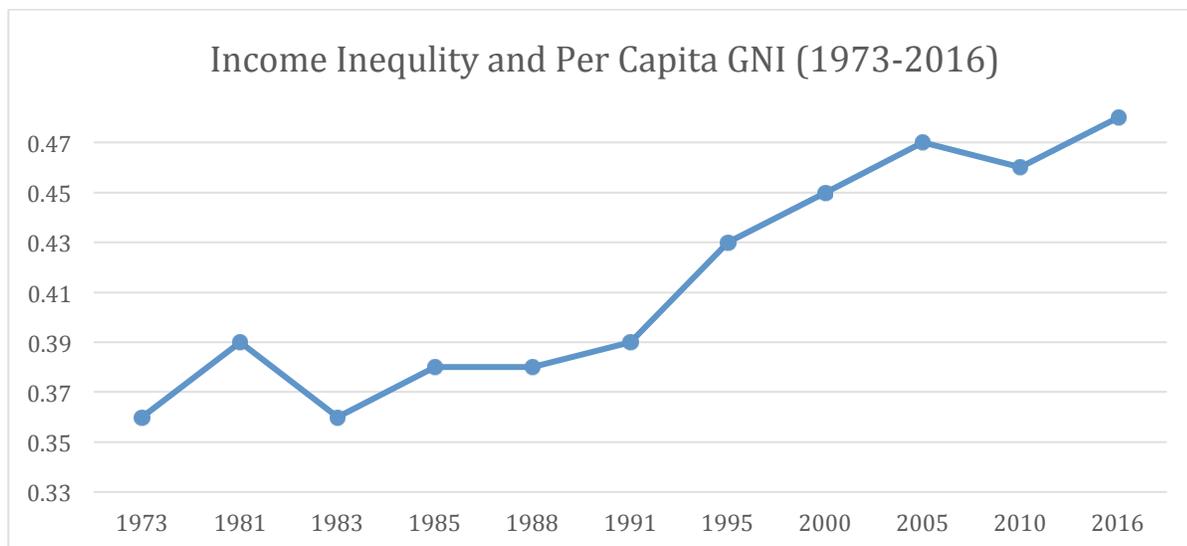

## X. CONCLUSIONS

The eradication of poverty and inequality are the primary goals for the sustainable development of any country. Bangladesh has achieved tremendous success to eradicate poverty and extreme poverty. Slower and unequal household consumption growth makes sloth the rate of poverty reduction. Average annual consumption fell from 1.8% to 1.4% from 2010 to 2016 and poorer households experienced slower consumption growth compared to richer households. In order to reduce poverty by an additional percentage point consumption growth should be distributed more equally. In Bangladesh poverty is much lower in urban areas than rural areas but poverty reduction in rural areas accounts for 90% of all poverty reduction that occurred between 2010 and 2016.

Karl Marx said income concentrated to top from bottom and inequality could increase over time. It is true for Bangladesh and it has been ranked the fastest growing country with an increasing number of rich population in the world. The number of ultra high net worth individuals in Bangladesh rose by 17.3% during the period where the United States rose by 8.1%. This indicates wealth and income concentrated to a richer portion of the country. Recent statistics show that the GINI Coefficient has been increasing over time in Bangladesh and income inequality is higher in urban areas.

## Bibliography

Zaman, K. A., & Akita, T. (201). Spatial Dimensions of Income Inequality and Poverty in Bangladesh: An Analysis of the 2005 and 2010 Household Income and Expenditure Survey Data. *Bangladesh Development Studies, XXXV* (3), 19-37.
Wodon, Q. T. *Growth, Poverty, and Inequality: A Regional Pale for Bangladesh.* The World Bank.
Wright, E. O., & Perrone, L. (1977). Marxist Class Categories and Income Distribution. *American Sociological Review, 42* (1), 32-55.
Islam, M. D., Sayeed, J., & Hossain, M. N. (2017). On Determinants of Poverty and Inequality in Bangladesh. *Journal of Poverty, 21* (4), 352-371.
Khan, M. A. Income Inequality in Bangladesh. *19th Biennial Conference "Rethinking Political Economy of Development.* Bangladesh Economic Association.